\pdfoutput=1

\documentclass[11pt]{article}

\usepackage{acl}

\usepackage{times}
\usepackage{latexsym}

\usepackage[T1]{fontenc}

\usepackage[utf8]{inputenc}

\usepackage{microtype}

\usepackage{hyperref}       %
\usepackage{url}            %
\usepackage{booktabs}       %
\usepackage{amsfonts}       %
\usepackage{nicefrac}       %
\usepackage{microtype}      %
\usepackage{xcolor}         %
\usepackage{amsfonts}
\usepackage{comment}
\usepackage{graphicx}
\usepackage{subcaption}
\usepackage{amsmath}
\usepackage{color}
\usepackage{lipsum}
\usepackage{hanging}
\usepackage{enumitem}
\usepackage{booktabs}
\usepackage{multirow}
\newcommand{\specialcell}[2][c]{%
  \begin{tabular}[#1]{@{}c@{}}#2\end{tabular}}
\newcommand{\eg}{\textit{e}.\textit{g}., }

\title{Revisiting Over-Smoothness in Text to Speech}

\author{
  Yi Ren
  \\
  Zhejiang University \\
  \texttt{rayeren@zju.edu.cn} \\\And
  Xu Tan  \\
  Microsoft Research \\
  \texttt{xuta@microsoft.com} \\\And
  Tao Qin \\ 
  Microsoft Research \\
  \texttt{taoqin@microsoft.com} \\\AND
  Zhou Zhao\thanks{\quad Corresponding author} \\
  Zhejiang University \\
  \texttt{zhaozhou@zju.edu.cn} \\\And
  Tie-Yan Liu \\
  Microsoft Research \\
  \texttt{tyliu@microsoft.com}
}

\begin{document}
\maketitle
\begin{abstract}
Non-autoregressive text to speech (NAR-TTS) models have attracted much attention from both academia and industry due to their fast generation speed. One limitation of NAR-TTS models is that they ignore the correlation in time and frequency domains while generating speech mel-spectrograms, and thus cause blurry and over-smoothed results. In this work, we revisit this over-smoothing problem from a novel perspective: the degree of over-smoothness is determined by the gap between the complexity of data distributions and the capability of modeling methods. Both simplifying data distributions and improving modeling methods can alleviate the problem. Accordingly, we first study methods reducing the complexity of data distributions. Then we conduct a comprehensive study on NAR-TTS models that use some advanced modeling methods. Based on these studies, we find that 1) methods that provide additional condition inputs reduce the complexity of data distributions to model, thus alleviating the over-smoothing problem and achieving better voice quality. 2) Among advanced modeling methods, Laplacian mixture loss performs well at modeling multimodal distributions and enjoys its simplicity, while GAN and Glow achieve the best voice quality while suffering from increased training or model complexity. 3) The two categories of methods can be combined to further alleviate the over-smoothness and improve the voice quality. 4) Our experiments on the multi-speaker dataset lead to similar conclusions as above and providing more variance information can reduce the difficulty of modeling the target data distribution and alleviate the requirements for model capacity.

\end{abstract}

\section{Introduction}
Non-autoregressive text to speech (NAR-TTS) models~\citep{ren2019fastspeech,ren2020fastspeech,peng2020non,vainer2020speedyspeech,lancucki2020fastpitch,kim2020glow,miao2020flow} have shown much faster inference speed than their autoregressive counterparts~\citep{wang2017tacotron,shen2018natural,ping2018deep}, while achieving comparable or even better voice quality~\citep{ren2019fastspeech,ren2020fastspeech}. The text-to-speech mapping can be formulated as a conditional distribution $P(y|x)$ where $x$ and $y$ are the text and speech sequences, respectively. Text-to-speech mapping is a one-to-many mapping problem~\citep{wang2017tacotron}, since multiple possible speech sequences correspond to a text sequence due to speech variations such as pitch, duration and prosody. Furthermore, speech mel-spectrograms are strongly correlated in time and frequency dimensions (see Section~\ref{sec:mel_dist} for detailed analyses). Therefore, $P(y|x)$ is actually a dependent and multimodal distribution~\citep{ling2013modeling,zen2014deep}\footnote{Here "dependent" means that the different dimensions (in either temporal domain or frequency domain) of $y$ are dependent to each other.}.

Early non-autoregressive TTS models~\citep{ren2019fastspeech,peng2020non} use mean absolute error (MAE) or mean square error (MSE) as loss function to model speech mel-spectrograms, implicitly assuming that data points in mel-spectrograms are independent to each other and follow a unimodal distribution\footnote{MAE can be derived from the Laplace distribution and MSE from the Gaussian distribution,  both of which are unimodal.}. Consequently, the mel-spectrograms following dependent and multimodal distributions cannot be well modeled by the MAE or MSE loss, which presents great challenges in non-autoregressive TTS modeling and causes over-smoothed (blurred) predictions in mel-spectrograms~\citep{vasquez2019melnet,sheng2019reducing}.

In this work, we conduct a comprehensive study on the over-smoothing problem in TTS. We find that the over-smoothness is closely related to the mismatch between the complexity of data distributions (\eg dependent and multimodal distributions are more complex than independent and unimodal distributions) and the power of modeling methods (\eg simple losses such as MAE and MSE are less powerful than GAN or Glow-based methods). Both simplifying data distributions and enhancing modeling methods can alleviate the over-smoothing problem. From this perspective, we categorize recent methods combating over-smoothness into two classes: \textbf{1. Simplifying data distributions}: The data distribution $P(y|x)$ can be simplified by providing more conditional input information. We review two main methods: 1) Providing the previous mel-spectrogram frames $y_{<t}$ to predict current frame $y_t$, i.e., factorizing the complex dependent distribution $P(y|x)$ into a simpler conditional distribution $\prod_t P(y_t|y_{<t}, x)$, as used in autoregressive TTS models~\citep{wang2017tacotron,li2018close}. 2) Providing more variance information $v$\footnote{The term "variance information" is first mentioned in FastSpeech 2~\citep{ren2020fastspeech}, which refers some speech-related conditional information} such as pitch, duration, and energy to predict mel-spectrogram in parallel, i.e., modeling $P(y|x, v)$ rather than $P(y|x)$, as done in some non-autoregressive TTS models~\citep{ren2020fastspeech,lancucki2020fastpitch}. \textbf{2. Enhancing modeling methods}: Generally speaking, the modeling method should be powerful enough to fit complex data distributions. We review methods based on different distribution assumptions, including Laplacian mixture loss, structural similarity index (SSIM)~\citep{wang2004image} loss, generative adversarial network (GAN)~\citep{lee2020multi} and Glow~\citep{kim2020glow}.

By studying those methods, we have the following findings. We hope that our studies and findings can inspire the community to design better models for TTS. 
\begin{itemize}[leftmargin=*]
\item By either autoregressive factorization or providing more variance information as input, complex distributions can be simplified to be less dependent and multimodal, which clearly alleviates the over-smoothing problem and improves the generated voice quality. Among them, providing more variance information such as FastSpeech 2 enjoys the advantages of fast generation due to its non-autoregressive nature.
\item Enhanced modeling methods outperform MAE in synthesized voice quality. Laplacian mixture loss significantly improves the quality of generated mel-spectrograms and enjoys the simplicity of modeling. GAN and Glow achieve the best quality under both subjective and objective evaluations (since they make no assumptions about output distributions), but at the cost of increased training or model complexity.

\item To further analyze the effectiveness of combining the basic ideas of the above two categories, we enhance FastSpeech 2 (considering its fast inference speed and good quality in the first category) with Laplacian mixture loss, SSIM, GAN and Glow, respectively. We find that the enhanced FastSpeech 2 generates speech with even better quality and the over-smoothing problem is further alleviated, which shows that the methods in the two categories are complementary to each other. 
\item We also extend our experiments to multi-speaker TTS task and obtain similar conclusions as above. Besides, we find that Glow has poor modeling ability in multi-speaker scenarios due to limited model capacity and more complex target data distributions compared with the single-speaker scenario, while it can be greatly alleviated by simplifying data distributions (introducing more variance information).
\end{itemize}

\section{Preliminary Study}

\begin{figure*}[!t]
	\centering
	\includegraphics[width=\textwidth,trim={0cm 0cm 0cm 0cm}, clip=true]{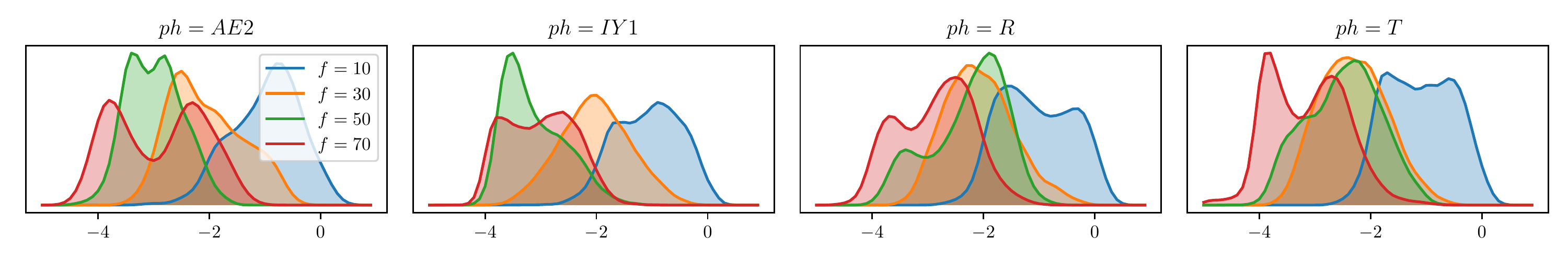}
	\vspace{-7mm}
	\caption{The marginal distributions $P(y(t,f)|x=ph)$ for several different phonemes.  We choose 4 different phonemes ($ph=AE2, IY1, R, T)$ and 4 frequency bins ($f=10, 20, 50, 70$) in this case study. More marginal distributions are added in Appendix \ref{sec:apdx_mel_dist}.}
	\label{fig:mel_margin}
\end{figure*}

\begin{figure*}[!t]
    \centering
    \begin{subfigure}{0.48\textwidth}
    	\centering
    	\includegraphics[width=\textwidth,trim={8cm 0cm 7.6cm 0cm}, clip=true]{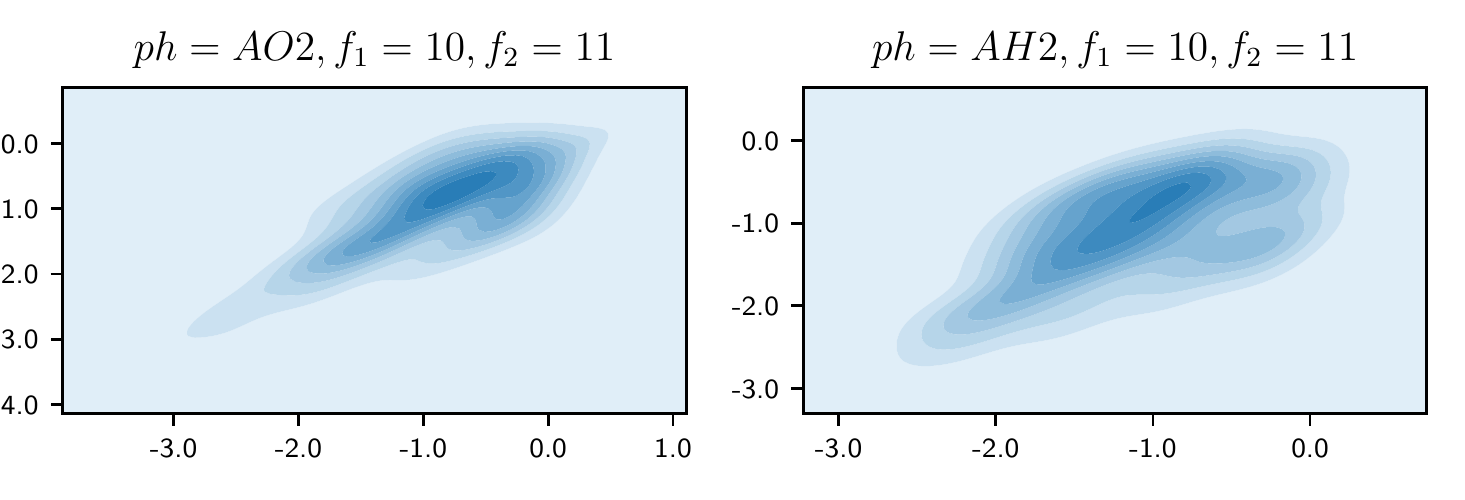}
    	\vspace{-7mm}
    	\caption{}
    	\label{fig:mel_joint_f}
    \end{subfigure}
    \begin{subfigure}{0.48\textwidth}
    	\centering
    	\includegraphics[width=\textwidth,trim={8cm 0cm 7.6cm 0cm}, clip=true]{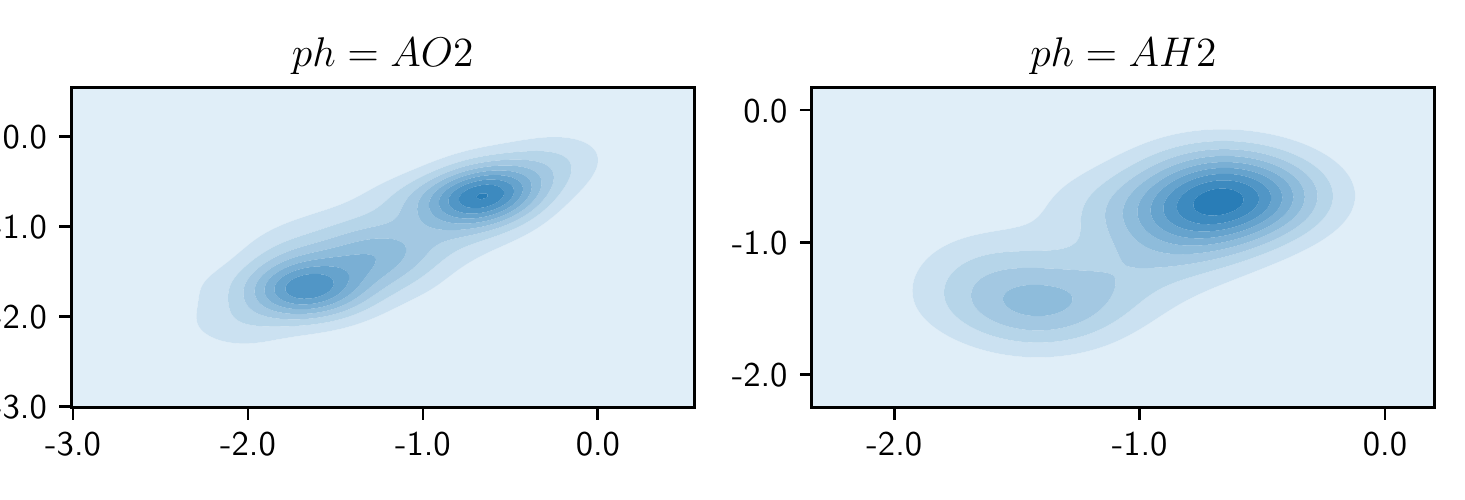}
    	\vspace{-7mm}
    	\caption{}
    	\label{fig:mel_joint_t}
    \end{subfigure}
    \caption{Joint distributions of two data points in mel-spectrogram. (a) Along the frequency axis $P(y(t,f_1), y(t,f_2)|x=ph)$, where $f_1=10$ and $f_2=11$. (b) Along the time axis $P(y(t_1,f), y(t_2,f)|x=ph)$. We choose 2 phonemes ($ph=AO2, AH2$) in this visualization. More visualizations are included in Appendix \ref{sec:apdx_mel_dist}.}
    \label{fig:mel_dist}
\end{figure*}

Text-to-speech mapping is a one-to-many mapping since multiple speech sequences can possibly correspond to a text sequence with different pitch, duration and prosody, making the mel-spectrograms distribution $P(y|x)$ multimodal. And due to the continuous nature of the mel-spectrograms, adjacent data points are dependent to each other. In this section, we first empirically characterize the distribution of $P(y|x)$ in TTS\footnote{In this work, we mainly focus on text (phoneme) to mel-spectrogram mapping, and leave mel-spectrogram to waveform mapping and text to waveform mapping to future work.} through visualization (Section \ref{sec:mel_dist}), and then provide a novel perspective to study the over-smoothing problem in TTS (Section \ref{sec:hypothesis}).

\subsection{Visualizations}
\label{sec:mel_dist}
We first visualize the distribution of $P(y|x)$ to see whether it is dependent and multimodal\footnote{We approximate the mel-spectrogram distribution on LJSpeech dataset. The detailed data processing procedure is the same as that in Section \ref{sec:exp_settings}.}. Denote the data point in the $t$-th frame and the $f$-th frequency bin in the ground-truth mel-spectrogram as $y(t,f)$, where $t \in [1, T]$, $f \in [1, F]$, and $T$, $F$ represent the total length and the number of frequency bins of mel-spectrograms respectively. Since different phonemes have different mel-spectrograms, we analyze the distribution of each phoneme separately. Specifically, for all the mel-spectrogram frames corresponding to each phoneme $ph$, we calculate three distributions: 1) marginal distribution $P(y(t,f)|x=ph)$; 2) joint distribution between two different frequency bins $P(y(t,f_1), y(t,f_2)|x=ph)$; 3) joint distribution between two different time frames $P(y(t_1,f), y(t_2,f)|x=ph)$\footnote{We denote one of the frame index among all mel-spectrograms corresponding to the phoneme $x$ as $t_1$ and set $t_2=t_1+1$.}. For each distribution, we first compute the histograms and smooth into probability density functions with kernel density estimation~\citep{dehnad1987density} for better visualization.

The marginal distributions $P(y(t,f)|x=ph)$ for several different phonemes are shown in Figure \ref{fig:mel_margin}. It can be seen that the shape of the marginal distribution of each data point $y(t,f)$ in mel-sepctrogram is multimodal, especially for data points in high-frequency bins. The joint distribution $P(y(t,f_1), y(t,f_2)|x=ph)$ and $P(y(t_1,f), y(t_2,f)|x=ph)$ are shown in Figure \ref{fig:mel_joint_f} and Figure \ref{fig:mel_joint_t}, respectively. Obviously, those joint distributions are also multimodal and neighboring points in mel-spectrograms are strongly correlated. From these observations, we can see that the distribution $P(y|x)$ of mel-sepctrograms is multimodal and dependent across time and frequency.

\subsection{A Novel Perspective}
\label{sec:hypothesis}

\begin{table}[!h]
    \small
	\centering
	\caption{The two categories of methods to combat over-smoothness in TTS.}
	\begin{tabular}{ c | c }
		\toprule
		Categories & Methods \\
		\midrule
		\specialcell{Simplifying \\ data distributions} & \specialcell{AR modeling \citep{li2018close}, \\ FastSpeech 2~\citep{ren2020fastspeech}, \\ FastPitch~\citep{lancucki2020fastpitch}} \\
		\midrule
		\specialcell{Enhancing \\ modeling methods} & \specialcell{
		SSIM~\citep{wang2004image}, \\
	    Laplacian mixture, \\
		GAN~\citep{lee2020multi}, \\
		Glow~\citep{kim2020glow}
		} \\
		\bottomrule
	\end{tabular}
	\label{tab:methods_cate}
\end{table}

The dependent and multimodal distribution of $P(y|x)$ increases the difficulty of TTS modeling and causes over-smoothing problem if it is not correctly handled. We provide a novel perspective to depict this problem: the degree of over-smoothness is closely related to the gap between the complexity of data distributions and the power of modeling methods. A larger gap between the power of a modeling method and the complexity of a data distribution results in more severe over-smoothing problem. Consequently, simplifying data distributions and enhancing modeling methods can alleviate the over-smoothing problem. From this perspective, we list the methods to combat over-smoothness in two categories in Table \ref{tab:methods_cate}. 

In the following sections, we first explore the effectiveness of simplifying data distributions (Section~\ref{sec:simplify_dist}) and then that of enhancing modeling methods (Section~\ref{sec:modeling_methods}). Finally, we combine the basic ideas of these two categories to improve an existing model (Section~\ref{sec:comparison}) and conduct further exploration on the multi-speaker dataset (Section~\ref{sec:multispk}).

\section{Simplifying Data Distributions $P(y|x)$}
\label{sec:simplify_dist}
Simplifying data distributions $P(y|x)$ is usually achieved by providing more conditional input information in the TTS literature~\citep{wang2017tacotron,li2018close,ren2020fastspeech}\footnote{\citet{ren2019fastspeech} use knowledge distillation to simplify the target mel-spectrogram itself, which also simplifies the data distribution but affects data quality as analyzed in ~\citet{ren2020fastspeech}. Thus, we do not consider this method in our study.}. In this way, more conditional information in input can alleviate the one-to-many mapping issue, and thus the distribution becomes less multimodal and the correlation along time and frequency is reduced given more condition. There are mainly two methods to provide more conditional input information: 1) autoregressive factorization along the time~\citep{wang2017tacotron,li2018close} or frequency axis; 2) providing more variance information to predict mel-spectrogram in parallel, as used in some non-autoregressive TTS models~\citep{ren2020fastspeech,lancucki2020fastpitch}. In this section, we first overview these two kinds of methods in detail and conduct the experiment analyses to measure their effectiveness in solving the over-smoothing problem.

\subsection{Methods}
In this subsection, we overview the two kinds of methods to simplify data distributions, including autoregressive factorization and providing more variance information as input.

\paragraph{Autoregressive Factorization}
The joint probability $P(y|x)$ can be factorized according to the chain rule in two ways along time and frequency dimensions respectively:
\begin{itemize}[leftmargin=*]
\item $P(y|x) = \prod_{t=1}^{T} P(y_t | y_{<t}, x)$, where $y_{<t}$ is the proceeding frames before the $t$-th frame and $T$ is the total frames. 
\item $P(y|x) = \prod_{f=1}^{F} P(y_f | y_{<f}, x)$, where $y_{<f}$ is the proceeding frequency bins before the $f$-th frequency bin and $F$ is the total number of frequency bins (\eg 80). 
\end{itemize}

\paragraph{More Variance Information}
Another way to simplify data distributions $P(y|x)$ is to provide more variance information $v$ such as pitch, duration, and energy, to convert $P(y|x)$ into $P(y|x, v)$, as used in previous works~\citep{ren2020fastspeech,lancucki2020fastpitch}. In this way, the distribution becomes less multimodal and the correlation along time and frequency is reduced.

\begin{figure}[!h]
	\centering
	\includegraphics[width=0.26\textwidth,trim={0cm 0cm 6cm 0cm}, clip=true]{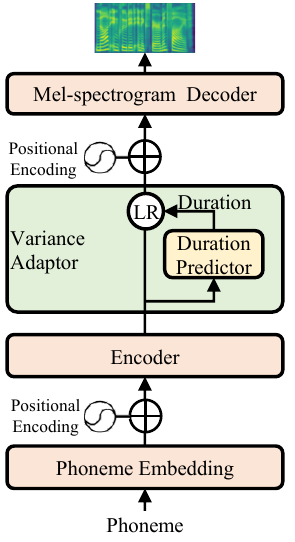}
	\caption{The overall architecture for our baseline model.}
	\label{fig:arch_fs1}
\end{figure}

\begin{figure*}[!t]
    \centering
    \newcommand{\MelDecoderVspace}{-2mm}
    \begin{subfigure}{0.26\textwidth}
    	\centering
    	\includegraphics[width=0.96\textwidth,trim={6.8cm 0.13cm 0.32cm 0.03cm}, clip=true]{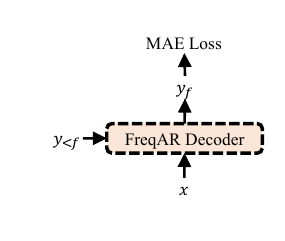}
    	\vspace{\MelDecoderVspace}
    	\caption{Frequency-AR.}
    	\label{fig:decoder_freqar}
    \end{subfigure}
    \begin{subfigure}{0.23\textwidth}
    	\centering
    	\includegraphics[width=0.8\textwidth,trim={2.65cm 0.1cm 5.2cm 0cm}, clip=true]{figs/arch_mel_decoders.pdf}
    	\vspace{\MelDecoderVspace}
    	\caption{Simple losses.}
    	\label{fig:decoder_simple}
    \end{subfigure}
    \begin{subfigure}{0.26\textwidth}
    	\centering
    	\includegraphics[width=0.88\textwidth,trim={0cm 0.10cm 7.28cm 0.01cm}, clip=true]{figs/arch_mel_decoders.pdf}
    	\vspace{\MelDecoderVspace}
    	\caption{GAN-based.}
    	\label{fig:decoder_gan}
    \end{subfigure}
    \begin{subfigure}{0.20\textwidth}
    	\centering
    	\includegraphics[width=0.85\textwidth,trim={4.85cm 0.15cm 3.2cm 0.05cm}, clip=true]{figs/arch_mel_decoders.pdf}
    	\vspace{\MelDecoderVspace}
    	\caption{Glow-based.}
    	\label{fig:decoder_glow}
    \end{subfigure}
    \caption{The architecture of mel-spectorgram decoder used in each mel-spectrogram modeling method. All these methods use the same architectures of phoneme embedding, encoder and variance adaptor as the baseline model in Figure \ref{fig:arch_fs1}. $x$ denotes the output hidden of the encoder.}
    \label{fig:mel_decoders}
\end{figure*}

\newcommand{\plotmel}[2]{
    \begin{subfigure}{0.185\textwidth}
	\centering
	\includegraphics[width=\textwidth,trim={2mm 2mm 2mm 2mm}, clip=true]{#1}
	\vspace{-6mm}
	\caption{\textit{#2}}
    \end{subfigure}
}
\begin{figure*}[!t]
    \centering
    \plotmel{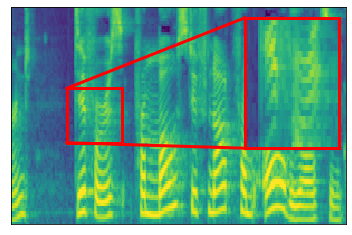}{GT}
    \plotmel{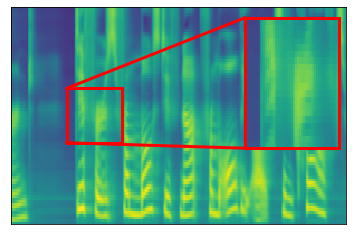}{MAE}
    \plotmel{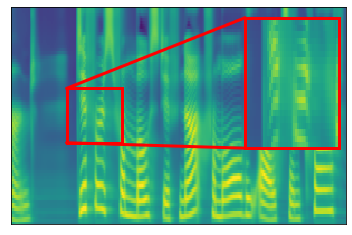}{FreqAR}
    \plotmel{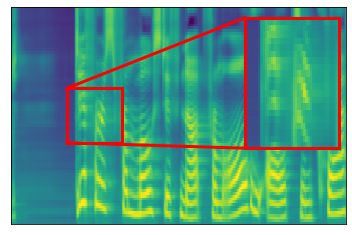}{TimeAR}
    \plotmel{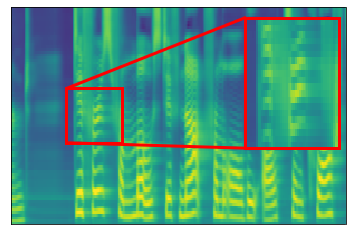}{FastSpeech 2}
    \caption{Visualizations of the ground-truth and generated mel-spectrograms by different methods for simplifying data distributions.}
    \label{fig:mel_main_vis1}
\end{figure*}

\subsection{Experiments and Analyses}
\paragraph{Experimental Settings} 
\label{sec:exp_settings}
We conduct all experiments\footnote{The corresponding audio samples are available at \url{https://revisittts.github.io/revisittts/}.} on LJSpeech dataset~\citep{ljspeech17}. We use ParallelWaveGAN (PWG)~\citep{yamamoto2020parallel} as the vocoder to convert mel-spectrograms to waveforms. To evaluate the voice quality of the synthesized speech subjectively, we conduct the MOS~\citep{loizou2011speech} tests. To measure the degree of over-smoothness of mel-spectrograms objectively, we calculate the variation of the Laplacian~\citep{pech2000diatom} ($\text{Var}_{\text{L}}$) on the generated mel-spectrograms and compare with that on the ground-truth mel-spectrograms. We use FastSpeech~\citep{ren2019fastspeech} trained with MAE loss as the baseline model as shown in Figure \ref{fig:arch_fs1}. For autoregressive modeling along time, we directly use TransformerTTS~\citep{li2018close}. For autoregressive modeling along frequency, we modify the vanilla mel-spectrogram decoder in baseline model to support autoregressive generation along frequency (which is called FreqAR decoder) and feed $y_{<f}$ to the FreqAR decoder to model $P(y_f | y_{<f})$ as shown in Figure \ref{fig:decoder_freqar}. For the method providing more variance information, we use FastSpeech 2~\citep{ren2020fastspeech}, which adds pitch, duration and energy information to the variance adaptor of the baseline model. We put more detailed model descriptions in Appendix \ref{sec:apdx_method_details} and experimental settings in Appendix \ref{sec:apdx_exp_settings}.

\paragraph{Results and Analyses} 

\begin{table}[!h]
\small
\centering
\caption{Results for different methods for simplifying data distributions for TTS. The best scores are in bold.}
\begin{tabular}{ l | c c c}
\toprule
Methods & MOS & $\text{Var}_{\text{L}}$ \\
\midrule
\textit{GT}          & 4.18$\pm$0.08 & 0.367 \\
\textit{GT (PWG)}    & 3.90$\pm$0.08 & /     \\
\midrule
\textit{MAE}         & 3.64$\pm$0.10 & 0.072 \\
\midrule
\textit{FreqAR}      & 3.80$\pm$0.09 & 0.175  \\
\textit{TimeAR}      & 3.81$\pm$0.08 & \textbf{0.186}  \\
\textit{FastSpeech 2}& \textbf{3.83$\pm$0.09} & 0.184  \\
\bottomrule
\end{tabular}
\label{tab:results_simplify_dist}
\end{table}

We conduct MOS evaluation and compute $\text{Var}_{\text{L}}$ to compare methods including the baseline model (denoted as \textit{MAE}), autoregressive modeling along frequency (denoted as \textit{FreqAR}) and along time (denoted as \textit{TimeAR}) and \textit{FastSpeech 2}. The results are shown in Table \ref{tab:results_simplify_dist}. We also visualize the mel-spectrograms generated by all modeling methods in Figure \ref{fig:mel_main_vis1}. From the results, we have some observations: 1) Autoregressive modeling along frequency (\textit{FreqAR}) and time (\textit{TimeAR}) dimensions both outperform \textit{MAE} in terms of MOS and $\text{Var}_{\text{L}}$, which shows that simplifying data distributions using frequency or time dimension factorization can ease the over-smoothing problem. However, the autoregressive models suffer from slow inference.
2) FastSpeech 2 also greatly outperforms the baseline model, further indicating that simplifying data distributions by providing more variance information is another way to alleviate the over-smoothing problem. 
In conclusion, autoregressive modeling and providing more variance information can both simplify the complex distribution to be less dependent and multimodal and thus alleviate the over-smoothing problem. Besides, methods that provide more variance information such as FastSpeech 2 also enjoy the fast inference speed.

\section{Enhancing Modeling Methods}
\label{sec:modeling_methods}
Most previous non-autoregressive TTS models~\citep{ren2019fastspeech,wang2019non,ren2020fastspeech} use mean absolute error (MAE) or mean square error (MSE) as training loss. However, they fail to capture dependent and multimodal distributions. MAE loss is derived from the Laplace distribution and MSE from the Gaussian distribution~\citep{chai2014root}, which means minimizing MAE/MSE will maximize the data log-likelihood under a Laplace/Gaussian distribution. Both of these distributions are unimodal and thus encourage the model to predict a single mode in each data point. As a result, the model just learns an average of all modes, which leads to over-smoothed results. Another problem brought by MAE and MSE is that they are independent across time and frequency for mel-spectrogram output, which ignores the correlation across time and frequency axes in mel-spectrogram.

In this section, we first introduce several enhanced modeling methods to directly model the dependent and multimodal distribution $P(y|x)$ (Section~\ref{sec:enhance_methods}), and then conduct experiments to compare and analyze these methods (Section~\ref{sec:enhance_exp}).

\subsection{Methods}
\label{sec:enhance_methods}
We list the enhanced methods, including SSIM loss, Laplacian mixture loss, GAN\footnote{Although GAN is well-known to suffer from the mode collapse issue, practically it performs very well in modeling the multi-modal distribution through well-tuning~\cite{mao2017least}. GAN can avoid the average frame prediction in MAE loss that has a strong unimodal assumption, and can generate high-quality and reasonable results when the data distribution is multimodal. Therefore, we regard GAN as "no distribution assumption" method.} and Glow-based method and their distribution assumptions in Table \ref{tab:dist_model}.  We put the details of each method in Appendix \ref{sec:apdx_method_details}.

\begin{table}[!h]
\small
\centering
\caption{The distribution assumptions of different methods.}
\begin{tabular}{ c | c }
	\toprule
	\textbf{Assumptions} &  \textbf{Modeling Methods} \\
	\midrule
	\midrule
	\specialcell{Independent \& \\ unimodal} & \specialcell{Mean absolute error} \\
	\midrule
	\specialcell{Independent \& \\ multimodal} & \specialcell{Laplacian mixture} \\
	\midrule
	No assumption  & \specialcell{
	SSIM~\cite{wang2004image}, \\ GAN~\cite{goodfellow2014generative}, \\ Glow~\cite{kingma2018glow}} \\
	\bottomrule
\end{tabular}
\label{tab:dist_model}
\end{table}

\paragraph{Structural Similarity Index (SSIM)}
Structural Similarity Index (SSIM)~\citep{wang2004image} is one of the state-of-the-art perceptual metrics to measure image quality, which can capture structural information and texture. The value of SSIM is between 0 and 1, where 1 indicates perfect perceptual quality relative to the ground truth. The model architecture of SSIM loss follows the baseline model in Figure \ref{fig:arch_fs1} and we directly replace the MAE loss in the baseline model with SSIM loss as shown in Figure \ref{fig:decoder_simple}. 

\paragraph{Laplacian Mixture (LM) Loss}
Laplacian mixture loss\footnote{We choose Laplace distribution as the mixture distribution since the distribution of the magnitude of spectrogram is Laplacian \citep{tits2019theory,usman2018probabilistic,gazor2003speech}. We have also tried other mixture distributions (\eg mixture of logistic and mixture of Gaussian) and have similar findings.} can model samples independently with multimodal distribution. As shown in Figure \ref{fig:decoder_simple}, the basic architecture of mel-spectrogram decoder follows baseline model and we modify the output layer of the baseline model to predict the multimodal distribution of each mel-spectrogram bin.

\paragraph{Generative Adversarial Network (GAN)}
We introduce adversarial training to better model the dependent and multimodal distribution. Inspired by \citet{wu2020adversarially,binkowski2019high}, we adopt multiple random window discriminators. We use the LSGAN~\citep{mao2017least} loss to train the TTS model and multi-window discriminators.

\paragraph{Glow}
Glow~\citep{kingma2018glow} is a kind of normalizing flow, which maps data into a known and simple prior (\eg spherical multivariate Gaussian distribution). As shown in Figure \ref{fig:decoder_glow}, our Glow-based decoder models the distribution of mel-spectrograms conditioned on the encoder output hidden states $x$.

\subsection{Experiments and Analyses}
\begin{figure}[!h]
\centering
\newcommand{\plotmelL}[2]{
    \begin{subfigure}{0.19\textwidth}
	\centering
	\includegraphics[width=\textwidth,trim={2mm 2mm 2mm 2mm}, clip=true]{#1}
	\caption{\textit{#2}}
    \end{subfigure}
}
\centering
\plotmelL{figs/mel_plots_main/GT.png}{GT}
\plotmelL{figs/mel_plots_main/0116_fs1_1.png}{MAE}
\plotmelL{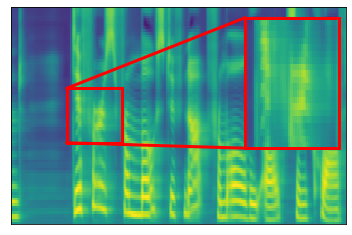}{SSIM}
\plotmelL{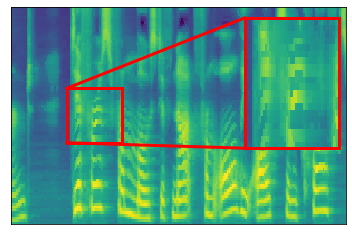}{LM}
\plotmelL{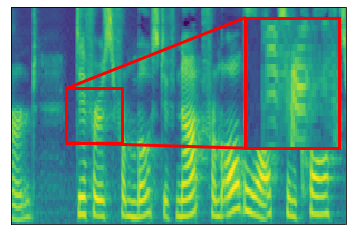}{GAN}
\plotmelL{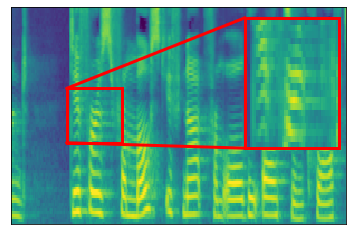}{Glow}
\caption{Visualizations of the ground-truth and generated mel-spectrograms by different modeling methods.}
\label{fig:mel_main_vis2}
\end{figure}

\label{sec:enhance_exp}
The dataset, baseline model and evaluation metrics are the same as Section \ref{sec:exp_settings}. We conduct MOS evaluation and compute $\text{Var}_{\text{L}}$ to compare different modeling methods including MAE (denoted as \textit{MAE}), Laplacian mixture loss (denoted as \textit{LM}), structural similarity index (denoted as \textit{SSIM}), generative adversarial network (denoted as \textit{GAN}), Glow (denoted as \textit{Glow}). The results are shown in Table \ref{tab:main_results_methods}. We also visualize mel-spectrograms generated by all modeling methods in Figure \ref{fig:mel_main_vis2}. From the results, we can find that:

\begin{table}[!h]
    \small
    \centering
    \caption{Results for different modeling methods for TTS. The best scores are in bold.}
    \begin{tabular}{l | c c c c}
    \toprule
    Methods & MOS & $\text{Var}_{\text{L}}$ \\
    \midrule
    \textit{GT}          & 4.18$\pm$0.08 & 0.367 \\
    \textit{GT (PWG)}    & 3.90$\pm$0.08 & /     \\
    \midrule
    \textit{MAE}         & 3.64$\pm$0.10 & 0.072 \\
    \midrule    
    \textit{SSIM}        & 3.68$\pm$0.10 & 0.201 \\
    \textit{LM}         & 3.72$\pm$0.08 & 0.296 \\
    \textit{GAN}         & 3.76$\pm$0.08 & \textbf{0.326} \\
    \textit{Glow}        & \textbf{3.78$\pm$0.08} & 0.310 \\
    \bottomrule
    \end{tabular}
    \label{tab:main_results_methods}
\end{table}

1) \textit{LM} and \textit{SSIM} outperform \textit{MAE} in terms of voice quality according to MOS evaluation and the mel-spectrogram visualizations, which shows that even simply replacing the loss function with those without strong unimodal and independent assumptions can significantly alleviate the over-smoothing problem and improve the generated mel-spectrograms. Among simple loss functions, \textit{LM} performs the best and generates sharper and clearer mel-spectrograms since its $\text{Var}_{\text{L}}$ is closer to \textit{GT}, which demonstrates that Laplacian mixture loss can model the multimodal distribution well. 

2) \textit{GAN} and \textit{Glow} show more superior performances compared with other modeling methods, indicating that modeling mel-spectrogram with both dependent and multimodal distribution can significantly ease the over-smoothing problem and improve the generated speech. From the visualizations, we can see that \textit{GAN} and \textit{Glow} generate formants with rich details in the middle/high-frequency region. However, GAN-based and Glow-based methods suffer from training complexity and large model footprint respectively: GAN relies on the discriminator to fit the distribution, which causes unstable training and difficult hyper-parameters tuning; Glow imposes strong architectural constraints and requires a large model footprint (about 2x model parameters) to keep the bijection between the simple independent latent distribution (\eg spherical multivariate Gaussian distribution) and the complex and dependent data distribution.

\section{Further Explorations and Extensions}
In this section, we first explore the advantages of combining methods from two categories and then perform extensional analyses on multi-speaker dataset. 

\subsection{Combining Methods from Two Categories}
\label{sec:comparison}
\begin{table}[!h]
\small
\centering
\caption{Results of different models that combine the basic ideas of two categories. The best scores are in bold.}
\begin{tabular}{ l | c | c c c c}
\toprule
Methods & MOS & $\text{Var}_{\text{L}}$ \\
\midrule
\textit{GT}                        & 4.27$\pm$0.10 & 0.367 \\
\textit{GT (PWG)}                  & 4.00$\pm$0.10 & /     \\
\midrule    
\textit{FastSpeech 2}              & 3.81$\pm$0.10 & 0.184  \\
\midrule
\textit{FastSpeech 2 + SSIM}       & 3.84$\pm$0.09 & 0.205 \\
\textit{FastSpeech 2 + LM}         & 3.90$\pm$0.10 & 0.306 \\
\textit{FastSpeech 2 + GAN}        & 3.86$\pm$0.10 & \textbf{0.341} \\
\textit{FastSpeech 2 + Glow}       & \textbf{3.92$\pm$0.10} & 0.315 \\
\bottomrule
\end{tabular}
\label{tab:results_comparison}
\end{table}

After studying the two categories to combat over-smoothing problems, we have demonstrated that both simplifying data distributions and enhancing modeling methods can alleviate this problem and improve the voice quality of TTS. A natural thought is to combine the methods from two categories, which may integrate the advantages of both aspects to reduce the gap between the complexity of data distribution and the power of the modeling method, resulting in better voice quality and further alleviating the over-smoothing problem. 
To demonstrate this idea, we choose the FastSpeech 2 as the model from the first category since it achieves better perceptual voice quality than autoregressive modeling models according to Table \ref{tab:results_simplify_dist} and importantly, it enjoys the fast and robust inference advantages due to its non-autoregressive nature. Then we combine two categories by applying enhanced modeling methods to FastSpeech 2 and obtain the following systems: 1) \textit{FastSpeech 2 + SSIM}, which replaces MAE loss with SSIM loss; 2) \textit{FastSpeech 2 + LM}, which predicts the k-component mixture of Laplace distribution and uses LM loss for training; 3) \textit{FastSpeech 2 + GAN}, which adds the adversarial loss to FastSpeech 2; and 4) \textit{FastSpeech 2 + Glow}, which replaces the mel-spectrogram decoder with Glow. We conduct subjective and objective evaluations to compare these combined systems with \textit{FastSpeech 2}.

The results are shown in Table \ref{tab:results_comparison}. We can see that combining FastSpeech 2 with more powerful modeling methods can further alleviate the over-smoothing problem in terms of $\text{Var}_{\text{L}}$ and improve the generated speech quality in terms of MOS, which demonstrate our idea that simplifying data distributions and enhancing modeling methods can be used together and it is complementary to each other to further improve TTS performance. Among these methods, \textit{GAN} performs best in alleviating the over-smoothing problem and \textit{Glow} achieves the best perceptual voice quality, while they suffer from the cost of increased training or model complexity as we described in Section \ref{sec:enhance_exp}; compared with \textit{GAN} and \textit{Glow}, \textit{LM} can generate mel-spectrograms with comparable clearness and quality while enjoying its simplicity.

\subsection{Analyses on Multi-Speaker Dataset}
\label{sec:multispk}

\begin{table}
\small
\centering
\caption{Results for of different multi-speaker TTS models on multi-speaker dataset. The best scores are in bold.}
\begin{tabular}{ l | c | c c c c}
\toprule
Methods & MOS & $\text{Var}_{\text{L}}$ \\
\midrule
\textit{GT}                        & 3.97$\pm$0.13 & 0.338 \\
\textit{GT (PWG)}                  & 3.67$\pm$0.10 & /     \\
\midrule    
\textit{FastSpeech}                & 2.96$\pm$0.10 & 0.050 \\
\textit{FastSpeech + GAN}          & 3.14$\pm$0.15 & 0.251 \\
\textit{FastSpeech + Glow}         & 2.94$\pm$0.10 & 0.232 \\
\textit{FastSpeech 2}              & 3.18$\pm$0.16 & 0.132 \\
\textit{FastSpeech 2 + GAN}        & \textbf{3.35$\pm$0.13} & 0.257 \\
\textit{FastSpeech 2 + Glow}       & 3.30$\pm$0.12 & \textbf{0.264} \\
\bottomrule
\end{tabular}
\label{tab:results_multispk}
\end{table}

To demonstrate the generalization of our findings and provide more insights, we conduct experiments on a multi-speaker LibriTTS~\citep{zen2019libritts} dataset. We modify our models to support multiple speakers by adding speaker embeddings to the encoder outputs to indicate the speaker identity. We put more details of our multi-speaker TTS models and the dataset in Appendix \ref{sec:apdx_multispk_settings}. We compare the following systems: 1) \textit{FastSpeech}; 2) \textit{FastSpeech + GAN}; 3) \textit{FastSpeech + Glow}; 4) \textit{FastSpeech 2}; 5) \textit{FastSpeech 2 + GAN}; and 6) \textit{FastSpeech 2 + Glow}. The results are shown in Table \ref{tab:results_multispk}. We can see that 1) simplifying data distributions by providing more variance information and enhancing modeling method can alleviate the over-smoothing problem and improve the generated mel-spectrogram, and combining them together can achieve further better audio quality, which are consistent with the findings on single-speaker dataset. 2) \textit{FastSpeech + Glow} leads to inferior performance compared with the baseline model (\textit{FastSpeech}), because the multi-speaker dataset has more complex target data distributions and Glow requires a large model footprint to capture them as described in Section \ref{sec:enhance_exp}. When providing more variance information, \textit{FastSpeech 2 + Glow} achieves much better performance, which can reduce the difficulty of modeling the target data distribution and thus alleviate the requirements for model capacity.

\section{Related Works}
\paragraph{Non-autoregressive Text to Speech}
Previous TTS systems such as Tacotron~\citep{wang2017tacotron}, Tacotron 2~\citep{shen2018natural}, Deep Voice 3~\citep{ping2018deep} and TransformerTTS~\citep{li2018close} synthesize speech sequentially, which suffer from slow inference speed. To solve these shortcomings, various non-autoregressive TTS models are proposed to synthesize spectrogram frames in parallel. FastSpeech~\citep{ren2019fastspeech} and ParaNet~\citep{peng2020non} are early non-autoregressive TTS model which both adopt a fully parallel model architecture and rely on an autoregressive teacher model to provide the alignment between phonemes and mel-spectrograms. FastSpeech introduces knowledge distillation for mel-spectrograms to simplify data distributions. FastSpeech 2~\citep{ren2020fastspeech} and FastPitch~\citep{lancucki2020fastpitch} introduce more variance information as input to further reduce the output uncertainty and ease the one-to-many mapping problem. However, they are trained with MAE loss, which fits independent and unimodal Laplace distribution and results in blurry and over-smoothed mel-spectrograms in inference. SpeedySpeech~\citep{vainer2020speedyspeech} use the combination of MAE and structural similarity index (SSIM) losses to avoid blurry mel-spectrograms. Glow-TTS~\citep{kim2020glow} and Flow-TTS~\citep{miao2020flow} both use flow-based decoder to apply some invertible transforms between mel-spectrograms and noise data sampled from simple distribution. \citet{sheng2019reducing} employ a cascaded Tacotron 2 and GAN pipeline to reduce the over-smoothness of synthesized speech. Multi-SpectroGAN~\citep{lee2020multi} introduces generative adversarial network (GAN) and a multi-scale discriminator and is trained with only the adversarial feedback by conditioning hidden states with variance information (\eg duration, pitch and energy) to a discriminator. GAN and Flow-based methods can model dependent and multimodal distribution well, while they suffer from training or model complexity. In this work, we conduct systematic studies on several modeling methods in both NAR-TTS and AR-TTS from a novel perspective.

\paragraph{Handling Dependent and Multimodal Distributions}
Dependent and multimodal distributions increase the uncertainty for model training and lead to blurry results, which is observed in many generation tasks \citep{gu2017non,isola2017image,mathieu2015deep}. There are some common ways to handle dependent and multimodal distributions: 1) using loss functions or modeling methods that can well fit the distributions; and 2) introducing some input variables or transforming the target data to simplify data distributions. In neural machine translation, \citet{gu2017non} tackle this problem by introducing knowledge distillation to simplify target data distributions and using fertilities extracted by an external aligner to directly model the nondeterminism in the translation process. In image translation task, \citet{isola2017image} compare the generated images by their proposed adversarial generative method with those generated by using MAE and MSE losses and conclude that adversarial loss can help avoid blurry results. In video prediction task, \citet{mathieu2015deep} propose a multi-scale architecture, an adversarial training method, and an image gradient difference loss function to deal with the inherently blurry predictions obtained from the MSE loss function. However, there is no systematic analysis on multimodal and dependent distributions in TTS task as far as we know. In this paper, we conduct comprehensive analyses and studies on handling dependent and multimodal distributions in TTS from a novel perspective.

\section{Conclusion}
In this paper, we revisited the over-smoothing problem in TTS with a novel perspective: the degree of over-smoothness is determined by the gap between the complexity of data distribution and the capability of the modeling method. Under this perspective, we classified existing methods combating over-smoothness into two categories: simplifying data distributions and enhancing modeling methods, and conducted comprehensive analyses and studies on these methods. For simplifying data distributions, we found that both AR factorization and providing more variance information as input (\eg FastSpeech 2) can alleviate the over-smoothing problem, and FastSpeech 2 enjoys the advantage of fast generation over AR factorization. For enhancing modeling methods, we found that Laplacian mixture loss can improve the generation quality and enjoy its simplicity, while GAN and Glow can further achieve better quality at the cost of increased training or model complexity. Based on the above findings, we further combined these two categories of methods and found that the over-smoothing problem is further alleviated, and the generated speech quality is further improved, which shows that these two categories are complementary to each other. When performing our analyses on the multi-speaker dataset, we drew similar conclusions and found that providing more variance information can reduce the difficulty of modeling the target data distribution and alleviate the requirements for model capacity. 

We hope that our studies can inspire the community and industry to develop more powerful TTS models. Besides, since we are the first to discuss the over-smoothing problem systematically in the speech domain. We hope our analysis methodology as well as the findings can be extended to other tasks and inspire other domains.

\bibliography{anthology,custom}

\begin{thebibliography}{41}
\expandafter\ifx\csname natexlab\endcsname\relax\def\natexlab#1{#1}\fi

\bibitem[{Arik et~al.(2017)Arik, Chrzanowski, Coates, Diamos, Gibiansky, Kang,
  Li, Miller, Ng, Raiman et~al.}]{arik2017deep}
Sercan~O Arik, Mike Chrzanowski, Adam Coates, Gregory Diamos, Andrew Gibiansky,
  Yongguo Kang, Xian Li, John Miller, Andrew Ng, Jonathan Raiman, et~al. 2017.
\newblock Deep voice: Real-time neural text-to-speech.
\newblock \emph{arXiv preprint arXiv:1702.07825}.

\bibitem[{Bi{\'n}kowski et~al.(2019)Bi{\'n}kowski, Donahue, Dieleman, Clark,
  Elsen, Casagrande, Cobo, and Simonyan}]{binkowski2019high}
Miko{\l}aj Bi{\'n}kowski, Jeff Donahue, Sander Dieleman, Aidan Clark, Erich
  Elsen, Norman Casagrande, Luis~C Cobo, and Karen Simonyan. 2019.
\newblock High fidelity speech synthesis with adversarial networks.
\newblock \emph{arXiv preprint arXiv:1909.11646}.

\bibitem[{Chai and Draxler(2014)}]{chai2014root}
Tianfeng Chai and Roland~R Draxler. 2014.
\newblock Root mean square error (rmse) or mean absolute error (mae)?
  --arguments against avoiding rmse in the literature.
\newblock \emph{Geoscientific model development}, 7(3):1247--1250.

\bibitem[{Dehnad(1987)}]{dehnad1987density}
Khosrow Dehnad. 1987.
\newblock Density estimation for statistics and data analysis.

\bibitem[{Gazor and Zhang(2003)}]{gazor2003speech}
Saeed Gazor and Wei Zhang. 2003.
\newblock Speech probability distribution.
\newblock \emph{IEEE Signal Processing Letters}, 10(7):204--207.

\bibitem[{Goodfellow et~al.(2014)Goodfellow, Pouget-Abadie, Mirza, Xu,
  Warde-Farley, Ozair, Courville, and Bengio}]{goodfellow2014generative}
Ian Goodfellow, Jean Pouget-Abadie, Mehdi Mirza, Bing Xu, David Warde-Farley,
  Sherjil Ozair, Aaron Courville, and Yoshua Bengio. 2014.
\newblock Generative adversarial nets.
\newblock In \emph{Advances in neural information processing systems}, pages
  2672--2680.

\bibitem[{Gu et~al.(2017)Gu, Bradbury, Xiong, Li, and Socher}]{gu2017non}
Jiatao Gu, James Bradbury, Caiming Xiong, Victor~OK Li, and Richard Socher.
  2017.
\newblock Non-autoregressive neural machine translation.
\newblock \emph{arXiv preprint arXiv:1711.02281}.

\bibitem[{Hartigan et~al.(1985)Hartigan, Hartigan et~al.}]{hartigan1985dip}
John~A Hartigan, Pamela~M Hartigan, et~al. 1985.
\newblock The dip test of unimodality.
\newblock \emph{Annals of statistics}, 13(1):70--84.

\bibitem[{Isola et~al.(2017)Isola, Zhu, Zhou, and Efros}]{isola2017image}
Phillip Isola, Jun-Yan Zhu, Tinghui Zhou, and Alexei~A Efros. 2017.
\newblock Image-to-image translation with conditional adversarial networks.
\newblock In \emph{Proceedings of the IEEE conference on computer vision and
  pattern recognition}, pages 1125--1134.

\bibitem[{Ito(2017)}]{ljspeech17}
Keith Ito. 2017.
\newblock The lj speech dataset.
\newblock \url{https://keithito.com/LJ-Speech-Dataset/}.

\bibitem[{Kim et~al.(2020)Kim, Kim, Kong, and Yoon}]{kim2020glow}
Jaehyeon Kim, Sungwon Kim, Jungil Kong, and Sungroh Yoon. 2020.
\newblock Glow-tts: A generative flow for text-to-speech via monotonic
  alignment search.
\newblock \emph{arXiv preprint arXiv:2005.11129}.

\bibitem[{Kingma and Ba(2014)}]{kingma2014adam}
Diederik~P Kingma and Jimmy Ba. 2014.
\newblock Adam: A method for stochastic optimization.
\newblock \emph{arXiv preprint arXiv:1412.6980}.

\bibitem[{Kingma and Dhariwal(2018)}]{kingma2018glow}
Durk~P Kingma and Prafulla Dhariwal. 2018.
\newblock Glow: Generative flow with invertible 1x1 convolutions.
\newblock In \emph{Advances in Neural Information Processing Systems}, pages
  10215--10224.

\bibitem[{{\L}a{\'n}cucki(2020)}]{lancucki2020fastpitch}
Adrian {\L}a{\'n}cucki. 2020.
\newblock Fastpitch: Parallel text-to-speech with pitch prediction.
\newblock \emph{arXiv preprint arXiv:2006.06873}.

\bibitem[{Lee et~al.(2020)Lee, Yoon, Noh, Kim, and Lee}]{lee2020multi}
Sang-Hoon Lee, Hyun-Wook Yoon, Hyeong-Rae Noh, Ji-Hoon Kim, and Seong-Whan Lee.
  2020.
\newblock Multi-spectrogan: High-diversity and high-fidelity spectrogram
  generation with adversarial style combination for speech synthesis.
\newblock \emph{arXiv preprint arXiv:2012.07267}.

\bibitem[{Li et~al.(2019)Li, Liu, Liu, Zhao, and Liu}]{li2018close}
Naihan Li, Shujie Liu, Yanqing Liu, Sheng Zhao, and Ming Liu. 2019.
\newblock Neural speech synthesis with transformer network.
\newblock In \emph{Proceedings of the AAAI Conference on Artificial
  Intelligence}, volume~33, pages 6706--6713.

\bibitem[{Ling et~al.(2013)Ling, Deng, and Yu}]{ling2013modeling}
Zhen-Hua Ling, Li~Deng, and Dong Yu. 2013.
\newblock Modeling spectral envelopes using restricted boltzmann machines and
  deep belief networks for statistical parametric speech synthesis.
\newblock \emph{IEEE transactions on audio, speech, and language processing},
  21(10):2129--2139.

\bibitem[{Loizou(2011)}]{loizou2011speech}
Philipos~C Loizou. 2011.
\newblock Speech quality assessment.
\newblock In \emph{Multimedia analysis, processing and communications}, pages
  623--654. Springer.

\bibitem[{Mao et~al.(2017)Mao, Li, Xie, Lau, Wang, and
  Paul~Smolley}]{mao2017least}
Xudong Mao, Qing Li, Haoran Xie, Raymond~YK Lau, Zhen Wang, and Stephen
  Paul~Smolley. 2017.
\newblock Least squares generative adversarial networks.
\newblock In \emph{Proceedings of the IEEE international conference on computer
  vision}, pages 2794--2802.

\bibitem[{Mathieu et~al.(2015)Mathieu, Couprie, and LeCun}]{mathieu2015deep}
Michael Mathieu, Camille Couprie, and Yann LeCun. 2015.
\newblock Deep multi-scale video prediction beyond mean square error.
\newblock \emph{arXiv preprint arXiv:1511.05440}.

\bibitem[{Miao et~al.(2020)Miao, Liang, Chen, Ma, Wang, and
  Xiao}]{miao2020flow}
Chenfeng Miao, Shuang Liang, Minchuan Chen, Jun Ma, Shaojun Wang, and Jing
  Xiao. 2020.
\newblock Flow-tts: A non-autoregressive network for text to speech based on
  flow.
\newblock In \emph{ICASSP 2020-2020 IEEE International Conference on Acoustics,
  Speech and Signal Processing (ICASSP)}, pages 7209--7213. IEEE.

\bibitem[{Pech-Pacheco et~al.(2000)Pech-Pacheco, Crist{\'o}bal,
  Chamorro-Martinez, and Fern{\'a}ndez-Valdivia}]{pech2000diatom}
Jos{\'e}~Luis Pech-Pacheco, Gabriel Crist{\'o}bal, Jes{\'u}s Chamorro-Martinez,
  and Joaqu{\'\i}n Fern{\'a}ndez-Valdivia. 2000.
\newblock Diatom autofocusing in brightfield microscopy: a comparative study.
\newblock In \emph{Proceedings 15th International Conference on Pattern
  Recognition. ICPR-2000}, volume~3, pages 314--317. IEEE.

\bibitem[{Peng et~al.(2020)Peng, Ping, Song, and Zhao}]{peng2020non}
Kainan Peng, Wei Ping, Zhao Song, and Kexin Zhao. 2020.
\newblock Non-autoregressive neural text-to-speech.
\newblock ICML.

\bibitem[{Ping et~al.(2018)Ping, Peng, Gibiansky, Arik, Kannan, Narang, Raiman,
  and Miller}]{ping2018deep}
Wei Ping, Kainan Peng, Andrew Gibiansky, Sercan~O. Arik, Ajay Kannan, Sharan
  Narang, Jonathan Raiman, and John Miller. 2018.
\newblock Deep voice 3: 2000-speaker neural text-to-speech.
\newblock In \emph{International Conference on Learning Representations}.

\bibitem[{Ren et~al.(2020)Ren, Hu, Qin, Zhao, Zhao, and
  Liu}]{ren2020fastspeech}
Yi~Ren, Chenxu Hu, Tao Qin, Sheng Zhao, Zhou Zhao, and Tie-Yan Liu. 2020.
\newblock Fastspeech 2: Fast and high-quality end-to-end text-to-speech.
\newblock \emph{arXiv preprint arXiv:2006.04558}.

\bibitem[{Ren et~al.(2019)Ren, Ruan, Tan, Qin, Zhao, Zhao, and
  Liu}]{ren2019fastspeech}
Yi~Ren, Yangjun Ruan, Xu~Tan, Tao Qin, Sheng Zhao, Zhou Zhao, and Tie-Yan Liu.
  2019.
\newblock Fastspeech: Fast, robust and controllable text to speech.
\newblock In \emph{Advances in Neural Information Processing Systems}, pages
  3165--3174.

\bibitem[{Shen et~al.(2018)Shen, Pang, Weiss, Schuster, Jaitly, Yang, Chen,
  Zhang, Wang, Skerrv-Ryan et~al.}]{shen2018natural}
Jonathan Shen, Ruoming Pang, Ron~J Weiss, Mike Schuster, Navdeep Jaitly,
  Zongheng Yang, Zhifeng Chen, Yu~Zhang, Yuxuan Wang, Rj~Skerrv-Ryan, et~al.
  2018.
\newblock Natural tts synthesis by conditioning wavenet on mel spectrogram
  predictions.
\newblock In \emph{2018 IEEE International Conference on Acoustics, Speech and
  Signal Processing (ICASSP)}, pages 4779--4783. IEEE.

\bibitem[{Sheng and Pavlovskiy(2019)}]{sheng2019reducing}
Leyuan Sheng and Evgeniy~N Pavlovskiy. 2019.
\newblock Reducing over-smoothness in speech synthesis using generative
  adversarial networks.
\newblock In \emph{2019 International Multi-Conference on Engineering, Computer
  and Information Sciences (SIBIRCON)}, pages 0972--0974. IEEE.

\bibitem[{Tits et~al.(2019)Tits, Haddad, and Dutoit}]{tits2019theory}
No{\'e} Tits, Kevin~El Haddad, and Thierry Dutoit. 2019.
\newblock The theory behind controllable expressive speech synthesis: a
  cross-disciplinary approach.
\newblock \emph{arXiv preprint arXiv:1910.06234}.

\bibitem[{Usman et~al.(2018)Usman, Zubair, Shiblee, Rodrigues, and
  Jaffar}]{usman2018probabilistic}
Mohammed Usman, Mohammed Zubair, Mohammad Shiblee, Paul Rodrigues, and Syed
  Jaffar. 2018.
\newblock Probabilistic modeling of speech in spectral domain using maximum
  likelihood estimation.
\newblock \emph{Symmetry}, 10(12):750.

\bibitem[{Vainer and Du{\v{s}}ek(2020)}]{vainer2020speedyspeech}
Jan Vainer and Ond{\v{r}}ej Du{\v{s}}ek. 2020.
\newblock Speedyspeech: Efficient neural speech synthesis.
\newblock \emph{arXiv preprint arXiv:2008.03802}.

\bibitem[{Van Den~Oord et~al.(2016)Van Den~Oord, Dieleman, Zen, Simonyan,
  Vinyals, Graves, Kalchbrenner, Senior, and Kavukcuoglu}]{van2016wavenet}
A{\"a}ron Van Den~Oord, Sander Dieleman, Heiga Zen, Karen Simonyan, Oriol
  Vinyals, Alex Graves, Nal Kalchbrenner, Andrew~W Senior, and Koray
  Kavukcuoglu. 2016.
\newblock Wavenet: A generative model for raw audio.
\newblock \emph{SSW}, 125.

\bibitem[{Vasquez and Lewis(2019)}]{vasquez2019melnet}
Sean Vasquez and Mike Lewis. 2019.
\newblock Melnet: A generative model for audio in the frequency domain.
\newblock \emph{arXiv preprint arXiv:1906.01083}.

\bibitem[{Vaswani et~al.(2017)Vaswani, Shazeer, Parmar, Uszkoreit, Jones,
  Gomez, Kaiser, and Polosukhin}]{vaswani2017attention}
Ashish Vaswani, Noam Shazeer, Niki Parmar, Jakob Uszkoreit, Llion Jones,
  Aidan~N Gomez, {\L}ukasz Kaiser, and Illia Polosukhin. 2017.
\newblock Attention is all you need.
\newblock In \emph{Advances in Neural Information Processing Systems}, pages
  5998--6008.

\bibitem[{Wang et~al.(2019)Wang, Tian, He, Qin, Zhai, and Liu}]{wang2019non}
Yiren Wang, Fei Tian, Di~He, Tao Qin, ChengXiang Zhai, and Tie-Yan Liu. 2019.
\newblock Non-autoregressive machine translation with auxiliary regularization.
\newblock In \emph{AAAI}.

\bibitem[{Wang et~al.(2017)Wang, Skerry-Ryan, Stanton, Wu, Weiss, Jaitly, Yang,
  Xiao, Chen, Bengio et~al.}]{wang2017tacotron}
Yuxuan Wang, RJ~Skerry-Ryan, Daisy Stanton, Yonghui Wu, Ron~J Weiss, Navdeep
  Jaitly, Zongheng Yang, Ying Xiao, Zhifeng Chen, Samy Bengio, et~al. 2017.
\newblock Tacotron: Towards end-to-end speech synthesis.
\newblock \emph{arXiv preprint arXiv:1703.10135}.

\bibitem[{Wang et~al.(2004)Wang, Bovik, Sheikh, and Simoncelli}]{wang2004image}
Zhou Wang, Alan~C Bovik, Hamid~R Sheikh, and Eero~P Simoncelli. 2004.
\newblock Image quality assessment: from error visibility to structural
  similarity.
\newblock \emph{IEEE transactions on image processing}, 13(4):600--612.

\bibitem[{Wu and Luan(2020)}]{wu2020adversarially}
Jie Wu and Jian Luan. 2020.
\newblock Adversarially trained multi-singer sequence-to-sequence singing
  synthesizer.
\newblock \emph{arXiv preprint arXiv:2006.10317}.

\bibitem[{Yamamoto et~al.(2020)Yamamoto, Song, and Kim}]{yamamoto2020parallel}
Ryuichi Yamamoto, Eunwoo Song, and Jae-Min Kim. 2020.
\newblock Parallel wavegan: A fast waveform generation model based on
  generative adversarial networks with multi-resolution spectrogram.
\newblock In \emph{ICASSP 2020-2020 IEEE International Conference on Acoustics,
  Speech and Signal Processing (ICASSP)}, pages 6199--6203. IEEE.

\bibitem[{Zen et~al.(2019)Zen, Dang, Clark, Zhang, Weiss, Jia, Chen, and
  Wu}]{zen2019libritts}
Heiga Zen, Viet Dang, Rob Clark, Yu~Zhang, Ron~J Weiss, Ye~Jia, Zhifeng Chen,
  and Yonghui Wu. 2019.
\newblock Libritts: A corpus derived from librispeech for text-to-speech.
\newblock \emph{arXiv preprint arXiv:1904.02882}.

\bibitem[{Zen and Senior(2014)}]{zen2014deep}
Heiga Zen and Andrew Senior. 2014.
\newblock Deep mixture density networks for acoustic modeling in statistical
  parametric speech synthesis.
\newblock In \emph{2014 IEEE international conference on acoustics, speech and
  signal processing (ICASSP)}, pages 3844--3848. IEEE.

\end{thebibliography}
\bibliographystyle{acl_natbib}

\appendix
\section{Details in Modeling Methods}
\label{sec:apdx_method_details}
\subsection{Baseline Model}
Our baseline model is based on FastSpeech~\cite{ren2019fastspeech}. The dimension of phoneme embeddings and the hidden size of the self-attention are set to 256; the number of attention heads is set to 2 and the kernel sizes of the 1D-convolution in the 2-layer convolutional network after the self-attention layer are set to 9 and 1, with input/output size of 256/1024 for the first layer and 1024/256 in the second layer; the output linear layer converts the 256-dimensional hidden states into 80-dimensional mel-spectrograms. The size of the phoneme vocabulary is 76, including punctuations. In the duration predictor, the kernel sizes of the 1D-convolution are set to 3, with input/output sizes of 256/256 for both layers and the dropout rate is set to 0.5.

\subsection{Autoregressive Modeling along Frequency}
\begin{figure}[!h]
	\centering
	\includegraphics[width=0.27\textwidth,trim={0cm 0cm 0cm 0cm}, clip=true]{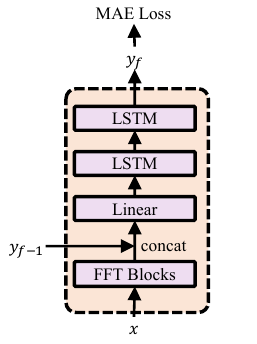}
	\caption{The architecture of the decoder for autoregressive modeling along frequency. ``FFT blocks" denotes the feedforward Transformer Blocks~\cite{ren2019fastspeech}.}
	\label{fig:freqar}
\end{figure}

Autoregressive modeling along frequency (denoted as FreqAR) factorizes $P(y|x)$ to $\prod_{f=1}^{F} P(y_f | y_{<f}, x)$, where $y_{<f}$ is the proceeding frequency bins before the $f$-th frequency bin and $F$ is the total number of frequency bins (\eg 80). We implement FreqAR by adding an extra small LSTM to the top of the four feedforward Transformer blocks\footnote{Since the time axis of mel-spectrogram has a variable length, we cannot model the autoregressive along frequency by 
regarding the time axis as the channel dimension and directly feeding the transposed mel-spectrogram into a causal decoder similar to that in TransformerTTS.} as shown in Figure \ref{fig:freqar}. For timestep $f=1$, we set the input hidden $h_0$ of LSTM to the output hidden of the mel-spectrogram decoder $q$, which has a size of $T \times H$, where $T$ is the total number of the frames, and $H$ is the hidden size of the LSTM. For $1<f<F$, we concatenate the previous output frequency bins $y_{f-1}$ with the size of $T \times 1$ and $q$ with the size of $T \times H$ along the channel axis and project the channel to $H$ with a linear layer as the input hidden $h_f$ of LSTM. The output hidden of LSTM is projected to $o_f$ using another linear layer. Finally, we concatenate $y_1, y_2, ..., y_{80}$ along the channel axis and obtain the output mel-spectrogram. We train the FreqAR model with teacher forcing. In this work, we set $H$ to $32$.

\subsection{Structural Similarity Index (SSIM)}
Structural Similarity Index (SSIM)~\citep{wang2004image} is one of the state-of-the-art perceptual metrics to measure image quality, which can capture structural information and texture. The value of SSIM is between 0 and 1, where 1 indicates perfect perceptual quality relative to the ground truth. For each data point in predicted and ground-truth mel-spectrograms ($\hat{s}=\hat{y}(t,f)$ and $s=y(t,f)$), the SSIM value is computed as:
$
\operatorname{SSIM}(s, \hat{s})=\frac{2 \mu_{s} \mu_{\hat{s}}+C_{1}}{\mu_{s}^{2}+\mu_{\hat{s}}^{2}+C_{1}} \cdot \frac{2 \sigma_{s \hat{s}}+C_{2}}{\sigma_{s}^{2}+\sigma_{\hat{s}}^{2}+C_{2}},
$
where $\mu_{s}$ and $\mu_{\hat{s}}$ denote the means for two regions, which are centered in $s$ and $\hat{s}$ within a 2D-window with size $(W, W)$ respectively, and we set $W$ to 11; $\sigma_{s}$ and $\sigma_{\hat{s}}$ are standard deviation for regions $s$ and $\hat{s}$; $\sigma_{s\hat{s}}$ is the covariance of regions $s$ and $\hat{s}$ and $C_{1}=0.0001$ and $C_{2}=0.0009$ are constant values for stabilizing the denominator. The SSIM loss for all mel-spectrogram data points $y(t,f)$ and $\hat{y}(t,f)$ is expressed as:
$
\mathcal{L}_{SSIM}=\frac{1}{TF}\sum_{t=1}^T\sum_{f=1}^F(1-SSIM(y(t,f), \hat{y}(t,f))).
$
The model architecture of SSIM loss follows the baseline model and we directly replace the MAE loss in the baseline model with SSIM loss as shown in Figure \ref{fig:decoder_simple}.

\subsection{Laplacian Mixture (LM) Loss}

Laplacian mixture loss\footnote{We choose Laplace distribution as the mixture distribution since the distribution of the magnitude of spectrogram is Laplacian \citep{tits2019theory,usman2018probabilistic,gazor2003speech}. We have also tried other mixture distributions (\eg mixture of logistic and mixture of Gaussian) and have similar findings.} can model samples independently with multimodal distribution. Let $\operatorname{La}(y;\mu, \beta)$ denotes the probability distribution function for a Laplace random variable $y$, where $\mu$ and $\beta$ are the mean and average absolute deviation (MAD) of the Laplace distribution. The $k$-component mixture of Laplace distribution is defined as follows:
$
P(y(t,f)) = \sum_{k=1}^K \pi_k(t,f) \operatorname{La}(y(t,f); \mu_k(t,f), \beta_k(t,f)),
$
where $\mu_k(t,f)$ and $\beta_k(t,f)$ are the mean and MAD of the distribution of the $k$-th Laplacian component. The log-likelihood loss under mixture of Laplace distribution is:
$
L_{LM} = - \frac{1}{TF}\sum_{t=1}^T\sum_{f=1}^F \log \left\{\sum_{k=1}^K \pi_k \operatorname{La}(y(t,f); \mu_k, \beta_k)\right\}.
$
As shown in Figure \ref{fig:decoder_simple}, the basic architecture of mel-spectrogram decoder follows baseline model and we modify the output layer of the baseline model to predict $\mu_k(t,k), \sigma_k(t,k)$ and $\pi_k(t,k)$ for each component k in each mel-spectrogram bin located in $(t,k)$. Then the model is optimized with $L_{LM}$ loss only and we set $k=5$. In inference, for each data point in the mel-spectrogram, we first choice the Laplacian component $k$ according to the probability $\pi_1, ..., \pi_K$ and then sample it from the Laplacian distribution $\operatorname{La}(y(t,f); \mu_k(t,f), \beta_k(t,f))$.

\subsection{Generative Adversarial Network (GAN)}
\label{sec:ana_advloss}
We introduce adversarial training to better model the dependent and multimodal distribution. Inspired by \citet{wu2020adversarially,binkowski2019high}, we adopt multiple random window discriminators where the input mel-spectrogram is randomly clipped into 3 clips with different window lengths, and each clip is fed into one discriminator. The discriminator consists of a 3-layer 2D-convolutional network\footnote{Through experiments, we find that the 2D-CNN-based network performs better than 1D-CNN in discriminator, especially for the reconstruction of high-frequency details.} with LeakyReLU activation, each followed by the batch normalization and dropout layer, and finally, an extra linear layer is added to project the hidden states to a probability to measure whether the input is a real mel-spectrogram sample. We use the LSGAN~\citep{mao2017least} loss to train the TTS model $G$ and multi-window discriminators $D_1$, $D_2$ and $D_3$:
$
L_{adv_D} = \sum_{i=1}^{3}\mathbb{E}_{y}(D_i(y)-1)^{2}+\mathbb{E}_{ \hat{y} }D_i(\hat{y})^{2},
L_{adv_G} = \frac{1}{3}\sum_{i=1}^{3}\mathbb{E}_{\hat{y}}(D_i(\hat{y})-1)^{2},
$
where $\hat{y}$ and $y$ are the generated and ground-truth mel-spectrograms.

\begin{figure}[!h]
	\centering
	\includegraphics[width=0.27\textwidth,trim={6.7cm 0cm 0cm 0.5cm}, clip=true]{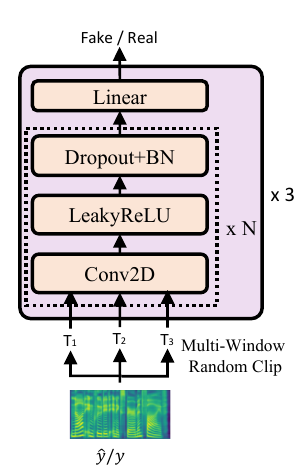}
	\caption{The architecture of multiple random window discriminators. $\hat{y}$ and $y$ denote the predicted and the ground-truth mel-spectrogram respectively.}
	\label{fig:disc}
\end{figure}

We use multiple random window discriminators to implement the generative adversarial network. The architecture of multiple random window discriminators is shown in Figure \ref{fig:disc}.

\subsection{Glow}
Glow~\citep{kingma2018glow} is a kind of normalizing flow, which maps data into a known and simple prior (\eg spherical multivariate Gaussian distribution). Glow is optimized with the exact log-likelihood of the data and as a generative model, it is very good at modeling all dependencies within very high-dimensional input data, which is usually specified in the form of a dependent and multimodal probability distribution. To make a fair comparison with other modeling methods, we only take the output hidden states of the encoder as the condition to each step of flows and use spherical Gaussian as the prior distribution like in the vanilla Glow~\citep{kingma2018glow}. As shown in Figure \ref{fig:decoder_glow}, our Glow-based decoder models the distribution of mel-spectrograms conditioned on the encoder output hidden states $x$. The Glow-based decoder is composed of $k$ flow steps $\boldsymbol{f}_1, ..., \boldsymbol{f}_k$, each of which consists of an affine coupling layer, an invertible 1x1 convolution, and an activation normalization layer. Glow-based decoder maps the spherical Gaussian random variables $\boldsymbol{z}$ to the mel-spectrograms as 
\begin{gather}
\boldsymbol{z} \sim \mathcal{N}(\boldsymbol{z};0,\boldsymbol{I}), 
~y = \boldsymbol{f}_0 \circ \boldsymbol{f}_1 \circ \ldots \boldsymbol{f}_k(\boldsymbol{z}, x).
\label{eq:glow_1}
\end{gather}

In training (dotted line in Figure \ref{fig:decoder_glow}), we directly minimize the negative log-likelihood of the data, which can be calculated using a change of variables: 
\begin{align}
\boldsymbol{z} = & \boldsymbol{f}_k^{-1} \circ \boldsymbol{f}_{k-1}^{-1} \circ \ldots \boldsymbol{f}_0^{-1}(y), \\
\log{p_\theta(y|x)} = & \log{p_\theta(\boldsymbol{z})} + \nonumber \\ & \sum_{i=1}^{k} \log
|\det(\boldsymbol{J}(\boldsymbol{f}_i^{-1}(y, x)))|,
\label{eq:glow_training_obj}
\end{align}
where the first term in Eq.~\eqref{eq:glow_training_obj} is the log-likelihood of the spherical Gaussian, $\boldsymbol{J}$ is the Jacobian and $\theta$ is the parameters of the Glow-based mel-spectrogram decoder. In inference (solid line in Figure \ref{fig:decoder_glow}), we sample $\boldsymbol{z}$ from the spherical multivariate Gaussian distribution and generate mel-spectrogram using Eq.~\eqref{eq:glow_1}.

\begin{figure}[!h]
	\centering
	\includegraphics[width=0.27\textwidth,trim={0cm 0cm 0cm 0cm}, clip=true]{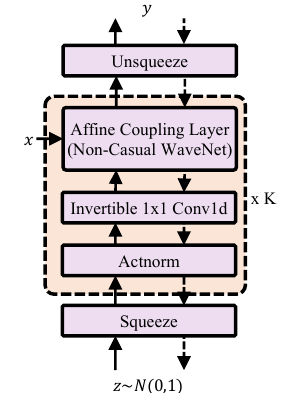}
	\caption{The architecture for Glow-based modeling method. The training and inference directions are represented with dotted and solid lines respectively.}
	\label{fig:glow}
\end{figure}

The architecture of Glow-based modeling method is shown in Figure \ref{fig:glow}. We choose non-causal WaveNet~\cite{van2016wavenet} as the network in the affine coupling layers following \citet{kim2020glow}. We set the number of flow steps $K$ to 24 and the layers of WaveNet to 4\footnote{The total number of parameters of Glow-based method (43M) is about twice that of other modeling methods (26M) and we find that the performance degrades when using the smaller model, indicating that the Glow-based modeling method requires a large model footprint to keep the bijection.}. We also set the dilation of WaveNet to 1 since we do not need very large receptive fields as that used in vocoder.

\begin{figure*}[!t]
	\centering
	\includegraphics[width=0.82\textwidth,trim={0cm 0cm 0cm 0cm}, clip=true]{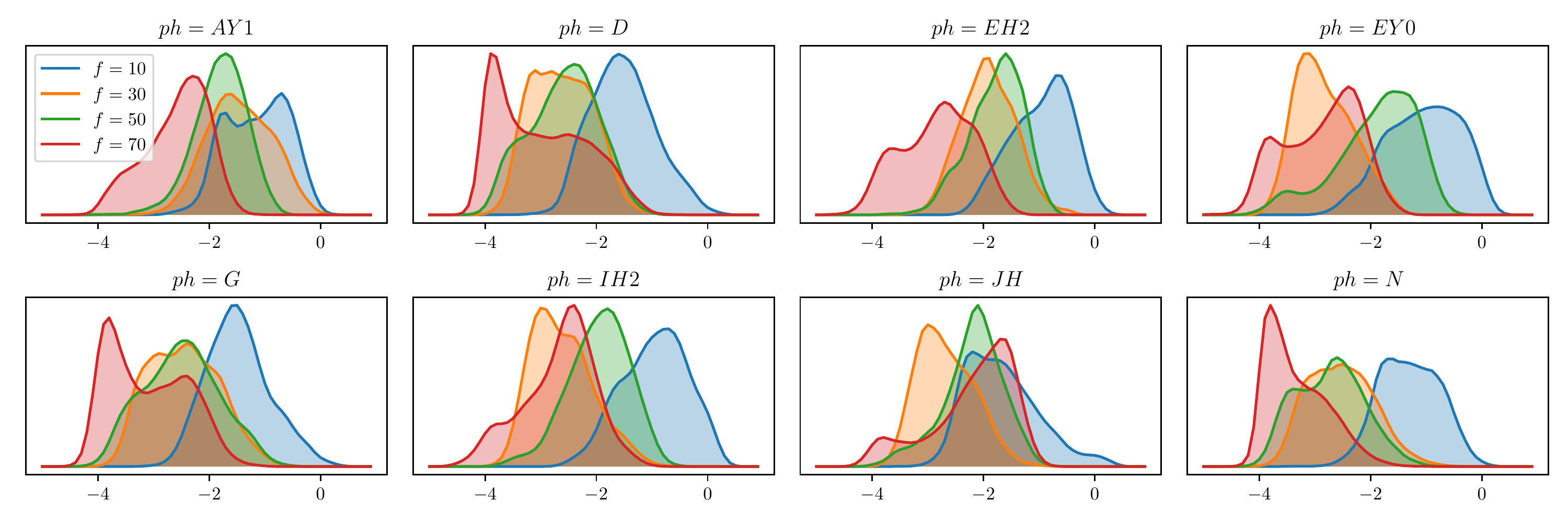}
	\caption{More marginal distributions mel-spectrogram $P(y(t,f)|x=ph)$ for several different phonemes $ph$.}
	\label{fig:mel_margin_all}
\end{figure*}

\begin{figure*}[!t]
	\centering
	\includegraphics[width=0.82\textwidth,trim={0cm 0cm 0cm 0cm}, clip=true]{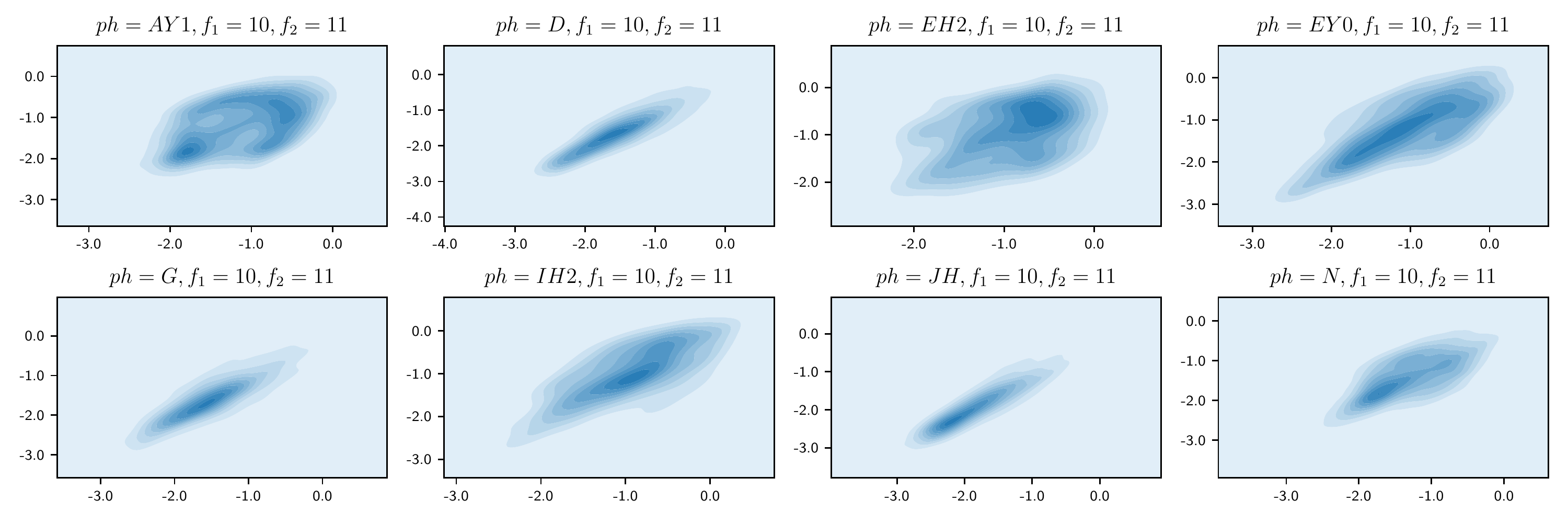}
	\caption{More joint distributions $P(y(t,f_1),y(t,f_2)|x)$ of two data points in mel-spectrogram along the frequency axis ($f_1=10$ and $f_2=11$).}
	\label{fig:mel_joint_all}
\end{figure*}

\begin{figure*}[!t]
	\centering
	\includegraphics[width=0.82\textwidth,trim={0cm 0cm 0cm 0cm}, clip=true]{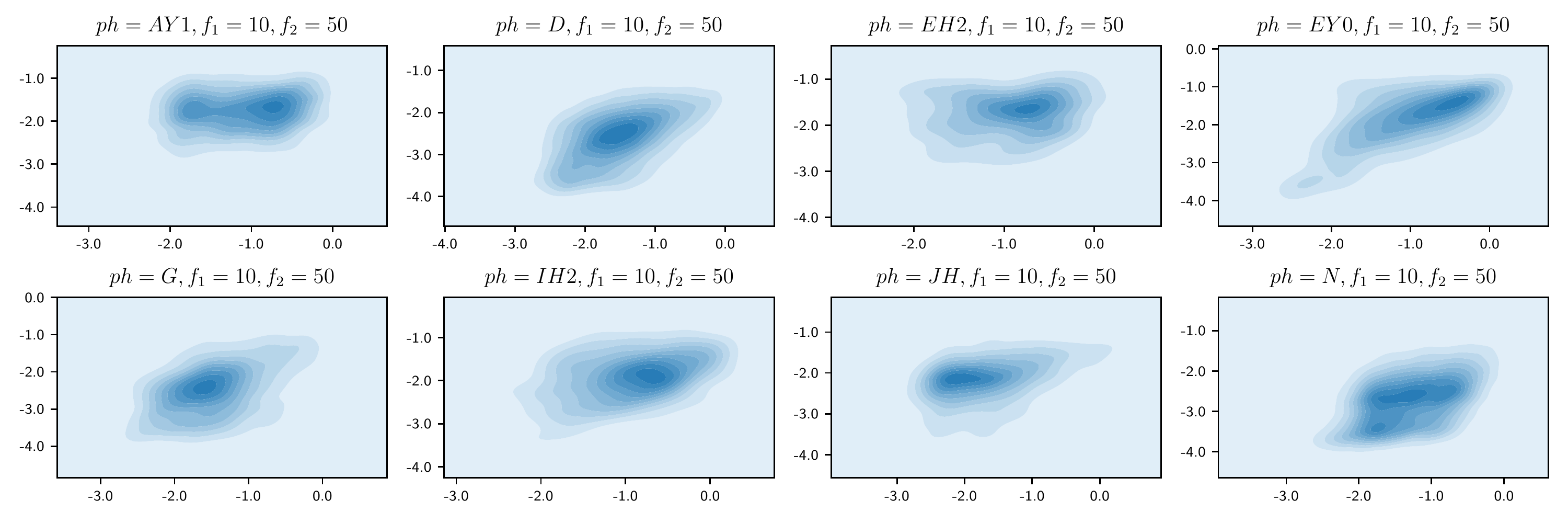}
	\caption{More joint distributions $P(y(t,f_1),y(t,f_2)|x)$ of two data points in mel-spectrogram along the frequency axis ($f_1=10$ and $f_2=50$).}
	\label{fig:mel_joint_far_all}
\end{figure*}

\begin{figure*}[!t]
	\centering
	\includegraphics[width=0.82\textwidth,trim={0cm 0cm 0cm 0cm}, clip=true]{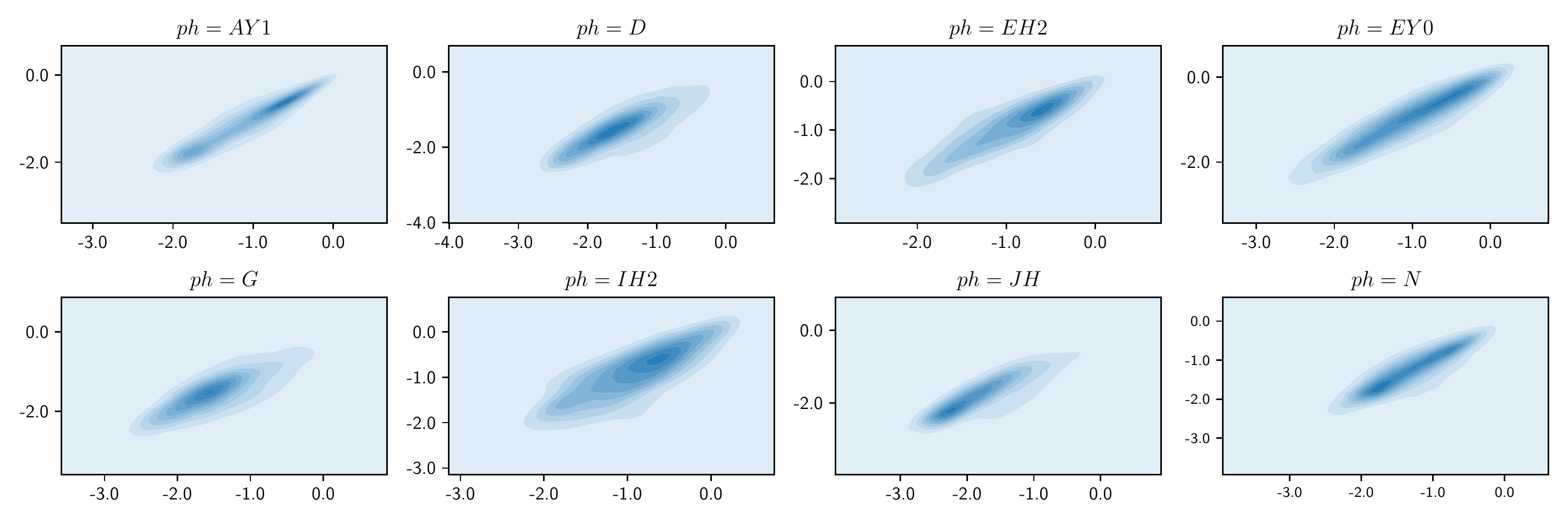}
	\caption{More joint distributions $P(y(t,f),y(t+1,f)|x)$ of two data points in mel-spectrogram along the time axis.}
	\label{fig:mel_joint_t_all}
\end{figure*}

\section{Experimental Settings} 
\label{sec:apdx_exp_settings}
In this section, we describe detailed experimental settings including datasets, training and inference details and evaluation criterion. 

\paragraph{Datasets}
We conduct all experiments on LJSpeech dataset~\citep{ljspeech17}, which contains 13,100 English audio clips (about 24 hours) of a single speaker and corresponding text transcripts. We split the dataset into three sets: 12,228 samples for training, 349 samples (with document title LJ003) for validation and 523 samples (with document title LJ001 and LJ002) for testing following ~\citet{ren2020fastspeech}. We convert the text sequence into the phoneme sequence~\citep{arik2017deep,wang2017tacotron,shen2018natural} with an open-source grapheme-to-phoneme tool\footnote{\url{https://github.com/Kyubyong/g2p}} to alleviate the mispronunciation problem. The raw waveform is transformed into mel-spectrograms following~\citet{shen2018natural} and we set frame size and hop size to 1024 and 256 with respect to the sample rate 22050.

\paragraph{Training and Inference}
\label{sec_exp_train_infer}
We train the model on 1 NVIDIA 2080Ti GPU, with batch size of 48 sentences. We use the Adam optimizer~\citep{kingma2014adam} with $\beta_{1}= 0.5$, $\beta_{2} = 0.998$ and $\varepsilon = 10^{-9}$ which can stabilize the adversarial training. We follow the same learning rate schedule in \citet{vaswani2017attention}. It takes 160k steps for training until convergence, except Glow-based model, which needs 480k steps. In the inference process, the output mel-spectrograms are transformed into audio samples using pre-trained Parallel WaveGAN~\citep{yamamoto2020parallel}\footnote{\url{https://github.com/kan-bayashi/ParallelWaveGAN}}. 

\paragraph{Subjective Evaluation}
To evaluate the perceptual quality, we conduct the MOS and CMOS~\citep{loizou2011speech} tests. We randomly choose 30 samples from the test set for subjective evaluation. Twenty native English speakers are asked to make quality judgments about the synthesized speech samples. The text content keeps consistent among different systems so that all testers only examine the audio quality without other interference factors. For MOS, each evaluator is asked to mark the subjective naturalness of a sentence on a 1-5 Likert scale. For CMOS, listeners are asked to compare pairs of audio generated by systems A and B and indicate which of the two audio they prefer and choose one of the following scores: 0 indicating no difference, 1 indicating small difference, 2 indicating a large difference and 3 indicating a very large difference. The estimated hourly wage paid to participants is about \$10 and we totally spent about \$900 on participant compensation.

\paragraph{Objective Evaluation}
\citet{pech2000diatom} propose the variation of the Laplacian as a ``blurriness metric" for images since the Laplacian filter can define edges well and blurry images barely have any edges while a well-focused (clear) image is expected to have a high variation of the Laplacian in grey levels. Inspired by their work, we introduce the variation of the Laplacian to mel-spectrograms and analyze its correlation with the smoothness. Specifically, the variation of the Laplacian $\text{Var}_{\text{L}}$ is given by:
$$
\text{Var}_{\text{L}}(\bar{y})=\sum_{t}^{T} \sum_{f}^{F}\left(|L(t, f)|-\frac{1}{NM}\sum_{m}^{M} \sum_{n}^{N}|L(m, n)|\right)^{2},
$$
where $L(t, f)$ is the convolution of the (predicted or ground-truth) mel-spectrogram $\bar{y}(t, f)$ with the Laplacian operation mask $L$ defined as:
$$
L=\frac{1}{6}\left(\begin{array}{rrr}
0 & -1 & 0 \\
-1 & 4 & -1 \\
0 & -1 & 0
\end{array}\right).
$$
The variation of the Laplacian increases with decreased smoothness of mel-spectrograms. We calculate the variation of the Laplacian $\text{Var}_{\text{L}}$ of the predicted mel-spectrogram and that of the ground-truth mel-spectrogram to see how close they are. We use an open-source tool\footnote{\url{https://github.com/petronav/Blur_Detection/blob/master/Variance_of_Laplacian/blur_check_vol.py}} to compute $\text{Var}_{\text{L}}$.

\section{Experimental Settings for Multi-Speaker TTS} 
\label{sec:apdx_multispk_settings}
In this section, we describe the experimental settings for multi-speaker TTS (Section 5.2). We first introduce the dataset, and then describe the model details.
\paragraph{Dataset}
We conduct our experiments on the \textit{train-clean-100} subset in LibriTTS~\citep{zen2019libritts}, which contains about 54 hours speech audio samples and their corresponding text transcriptions. We choose the \textit{train-clean-100} subset since the data in this subset is clean and the total time of speech audio in this subset is comparable with that in LJSpeech, which help us analyze the influence of multi-speaker setting while excluding the influence of training data size and noisy audio. These audio samples are recorded by 123 female speakers and 124 male speakers. We randomly choose 200 audio samples as the validation set, 200 of them as the test set and the rest of them as the training set. 
\paragraph{Model Details}
Based on the single-speaker baseline model described in Section 3.2, to introduce the speaker identity information into our model, we add an extra speaker embedding module. We look up the speaker embedding from this module and add it to the encoder outputs. The hidden size of the speaker embedding module is the same as the encoder hidden size.

\section{More Analyses on Mel-Spectrogram Distributions}
\subsection{More Visualizations}
\label{sec:apdx_mel_dist}
We plot more marginal and joint distributions (along time and frequency) of mel-spectrogram in Figure \ref{fig:mel_margin_all}, \ref{fig:mel_joint_all}, \ref{fig:mel_joint_far_all} and \ref{fig:mel_joint_t_all}. 

\subsection{Multimodality Evaluation on the Marginal Distributions}
\label{sec:apdx_mm_eval_margin}
To further demonstrate that combination of the methods from two categories (FastSpeech 2 and other enhanced modeling methods) can alleviate the over-smoothing problem, we conduct Hartigan's dip test~\cite{hartigan1985dip} on the marginal distributions $P(y(t,f)|x=ph)$ given $f$ and $ph$ to measure the degree of multimodality of each method. Specifically, we denote the dip test value given the distribution $P$ as $\mathcal{D}(P)$ and a lower value means more multimodal. we calculate the $\mathcal{D(P)}$ under the marginal distributions $P(\bar{y}(t,f)|x=ph)$ and average $\mathcal{D(P)}$ under different $ph$ and $f$ to obtain the averaged dip test value $\bar{\mathcal{D}}$. We compute $\bar{\mathcal{D}}$ on the following systems: 1) \textit{GT}, the ground-truth mel-spectrogram; 2) \textit{MAE}, which is the baseline model as described in Section 3.2; 3) \textit{FastSpeech 2} as described in Section 3.1;  4) \textit{FastSpeech 2 + SSIM}, which replaces MAE loss with SSIM loss; 5) \textit{FastSpeech 2 + LM}, which predicts the k-component mixture of Laplace distribution and is trained with LM loss; 6) \textit{FastSpeech 2 + GAN}, which adds the adversarial loss to FastSpeech 2; and 7) \textit{FastSpeech 2 + Glow}, which replaces the mel-spectrogram decoder with Glow.

\begin{table}[!h]
\small
\centering
\caption{Results of different models combining the basic ideas of two categories. The best scores are in bold.}
\vspace{2mm}
\begin{tabular}{ l | c}
\toprule
Methods & $\bar{\mathcal{D}}$ \\
\midrule
\textit{GT}                         & 0.049 \\
\textit{MAE}                        & 0.064  \\
\midrule    
\textit{FastSpeech 2}               & 0.060  \\
\textit{FastSpeech 2 + SSIM}        & 0.058 \\
\textit{FastSpeech 2 + LM}          & 0.060 \\
\textit{FastSpeech 2 + GAN}         & \textbf{0.054} \\
\textit{FastSpeech 2 + Glow}        & 0.056 \\
\bottomrule
\end{tabular}
\label{tab:results_dip}
\end{table}

The results are shown in Table \ref{tab:results_dip}. We can see that all of these combined methods can increase the degree of multimodality of the marginal distributions $P(y(t,f)|x=ph)$ and GAN achieves the best $\bar{\mathcal{D}}$ among these methods, which is consistent with our findings in Section 5 that all the combined methods can alleviate the over-smoothing problem and GAN performs the best.

\section{Potential Negative Societal Impacts} 
Although our analyses can inspire the community and industry to develop more powerful TTS models, it may result in unemployment for people with related occupations such as broadcaster and radio host. Besides, powerful TTS systems may be used in non-consensual voice cloning and fake media generation, which might be harmful to society.

\end{document}